\documentclass[pra,10pt,twocolumn,superscriptaddress]{revtex4-1}

\usepackage{epsfig}
\usepackage{amsmath}
\usepackage{graphicx}
\usepackage{graphics}
\usepackage{float}
\usepackage{amssymb}
\usepackage[usenames,dvipsnames]{color}
\usepackage{natbib}
\usepackage{epstopdf}
\usepackage{verbatim}

\newcommand{\bra}[1]{\left\langle #1\right|}
\newcommand{\ket}[1]{\left| #1\right\rangle}

\begin{document}

\title{Decoherence control: Universal protection of two-qubit states and two-qubit gates using continuous driving fields}

\author{Adam Zaman Chaudhry}
\affiliation{NUS Graduate School for Integrative Sciences
and Engineering, Singapore 117597, Singapore}
\author{Jiangbin Gong}
\email{phygj@nus.edu.sg}
\affiliation{NUS Graduate School for Integrative Sciences
and Engineering, Singapore 117597, Singapore}
\affiliation{Department of Physics and Center for Computational
Science and Engineering, National University of Singapore, 117542,
Singapore}

\begin{abstract}
A field configuration utilizing local static and oscillating fields is constructed to achieve universal (but low-order)
protection of two-qubit states.
That is,
two-qubit states can be protected against arbitrary system-environment coupling with a driving field whose frequency  is sufficiently large
as compared with the cutoff frequency of the environment.  Equally important,
we show that it is possible to construct driving fields to protect two-qubit entangling gates against decoherence, without assuming any particular form of system-environment coupling.   Using a non-Markovian master equation, we further demonstrate the effectiveness of our continuous dynamical decoupling fields in protecting entanglement and the excellent performance of protected two-qubit gates in generating entanglement. The  results are complementary to current
studies of entanglement protection using universal dynamical decoupling pulse sequences.
\end{abstract}

\maketitle

\section{INTRODUCTION}

The unwanted interaction between a system and its environment causes decoherence, i.e., the loss of quantum coherence.
Since future quantum technologies rely on coherent quantum states, it has become increasingly important to effectively
suppress decoherence.
To achieve this goal, various schemes, such as error-correction codes \cite{SteanePRL1996}, decoherence-free subspaces \cite{BaconPRA1999,ZanardiPRA1998} and dynamical decoupling (DD) \cite{Viola1998,LloydPRL1999,LloydPRL2000,LidarPRL2002,LiuFrontiers2011}, have been proposed. It is expected that in a large-scale working quantum computer, all of these schemes will be used in some way to store quantum states with high fidelity. For instance, a combination of dynamical decoupling and quantum error-correction codes has been proposed to combat errors due to spontaneous emission \cite{LidarPRA2003} (see also Ref.~\cite{KhoonPRA2011} for a recent study on such a hybrid scheme).

Our focus here is on DD. In the DD approach, external time-dependent fields are applied to the system such that the interaction term between the system and its environment rapidly flips sign. In this way, the effect of the environment on the system is canceled to a certain degree. The key advantage of DD, compared to some other methods such as quantum error-correction codes, is that no overhead is required - the qubits storing quantum information are protected directly, without any need for extra qubits.
Moreover, DD requires neither quantum measurements nor feedback control.

Broadly speaking, two types of DD have been studied: pulsed DD, which uses sharp pulses (impulsive pulses in many cases) to counter the effect of the environment, and continuous DD, which uses continuous-wave driving, that is, fields with simple harmonic time dependence.  We discuss pulsed DD first. The pioneering
 work on pulsed DD considered pulses applied, at equal time intervals, to a single qubit coupled to an environment \cite{Viola1998}.  Many extensions
 have been worked out since then, including, for example, DD for an arbitrary finite-dimensional system \cite{LloydPRL1999}
 and the suppression of arbitrary internal coupling in a quantum register \cite{Mahler}.  Studies of non-impulsive DD are also fruitful, by use of, for example, the so-called ``Eulerian DD" schemes~\cite{violaPRL2003,wocjanPRA2006}, and optimized pulses under an energy cost constraint \cite{KurizkiPRL2008} or a minimum leakage requirement \cite{WuPRL2007}.
 Recently, a significant advance was made when it was shown by Uhrig \cite{UhrigPRL2007} that by using aperiodic pulses, the so-called Uhrig's DD (UDD) scheme, the coherence of a single qubit can be protected to the $N^{\mbox{\scriptsize th}}$ order by using only $N$ (or $N+1$) instantaneous pulses. Uhrig originally considered only pure dephasing in the spin-boson model. Yang and Liu \cite{LiuPRL2008} then showed that UDD is universal in the sense that it does not depend on how a single qubit is coupled to its environment. Going further, West \emph{el al.}\:\cite{WestPRL2010} constructed a nested UDD sequence that can protect a single qubit against both dephasing and relaxation at the same time. A
mathematical proof for the effectiveness of nested UDD sequences has been recently given in Ref.\:\cite{KuoarXiv2011}. Effects of nonideal pulses on UDD are also under investigation~\cite{UhrigEPL2011}. On the experimental side, the excellent performance of UDD in protecting single-qubit quantum states has been studied in Refs.\:\cite{BiercukNature2009,BiercukPRA2009,LiuNature2009}.

Given high-efficiency single-qubit DD schemes,  extending single-qubit DD to two-qubit (or multi-qubit) decoherence control becomes more interesting.
It should be emphasized at this point that two-qubit (multi-qubit) decoherence control offers a whole new set of challenges (see also \cite{LloydPRL1999,Mahler,wocjanPRA2006}).
For instance, a fundamental objective of two-qubit decoherence control must be to protect two-qubit entanglement, since entanglement has been identified as the key resource for quantum information \cite{NielsenChuang}. As is now well known, quantum entanglement can, unlike single-qubit coherence, vanish in a finite amount of time \cite{HorodeckiPRA2001,EberlyPRL2004,FicekPRA2006,SainzPRA2008,EberlyScience2009,AlmeidaScience2007}.  In addition, there are different types of system-environment coupling that are not present in
single-qubit cases. In particular, there can be a noisy interaction between two qubits, and errors such as \textit{correlated} bit flipping, dissipation, and dephasing might emerge. Along this general direction of two-qubit DD, Ref.\:\cite{MukhtarPRA12010} showed for the first time that it is possible to
construct a pulse sequence to protect a known two-qubit quantum state to the $N$th order using $N$ pulses,
without any knowledge of system-environment coupling.  During the same year, the same group
of authors advocated the use of a nested sequence of UDD pulses to protect unknown two-qubit states with high efficiency, with each layer eliminating different noise terms \cite{MukhtarPRA22010}. It is now clear that to protect a completely unknown two-qubit state with high efficiency,
four layers of UDD pulses are required \cite{MukhtarPRA22010,LiuPRA2011}. These schemes are exciting because they are universal. That is,
so long as the pulses are applied fast enough (as compared with the cutoff frequency of the environment), we do not need to assume anything about the actual form of the system-environment coupling. Nested-UDD schemes have also been extended to multi-qubit systems with remarkable mathematical insights \cite{LiuPRA2011,LiangJiangarXiv2011}.  Parallel with these theoretical advances, preliminary experiments on entanglement protection using pulsed DD
have been performed in Refs. \cite{WangPRL2011,RoyPRA2011,ShuklaarXiv2011}.

Interestingly, many pulsed DD schemes mentioned above can be considered to be too strong in the sense that, while it does protect a quantum state with high efficiency, it also generally freezes useful coherent evolution generated by the system's own Hamiltonian.  To achieve useful coherent evolution concurrently with pulsed DD \cite{LloydPRL21999, LidarPRA2008}, one idea is to encode the logical qubits in physical qubits and then design the fields in such a way that the gate operation commutes with the pulse operations \cite{LidarPRL2010}. However, apparently an overhead is required.
Recently, a general procedure utilizing finite-power and finite-bandwidth pulses has been worked out for constructing dynamically corrected gates (DCG) without encoding or measurement overhead \cite{violaPRL2009,violaPRA2009}. Going further, by concatenating DCG's, it is possible to achieve arbitrary accuracy in quantum gate implementation \cite{violaPRL2010}. However, for arbitrary system-environment coupling,
dynamically corrected two-qubit gates have not been explicitly constructed and it is unclear how complicated the solution might be.

The existence of universal UDD schemes for two-qubit entanglement protection motivated this work.  In particular,
we shall investigate the usefulness of universal continuous DD (more specifically, DD based on driving fields with simple harmonic time dependence)
to protect two-qubit states, and hopefully, also to protect two-qubit gates.
Switching from pulsed DD to continuous DD, the sacrifice is obvious as compared with UDD and nested-UDD schemes: the performance of continuous DD is of a low-order nature. But our interest here is not with the high-order performance of a DD scheme. Rather,
we ask the following important question: Are there universal continuous DD schemes to protect two-qubit states and two-qubit gates, irrespective of how a two-qubit system is coupled with its environment?
This is a pertinent question to ask because, compared with pulsed DD, continuous DD has some advantages from a practical point of view.
For example, there is no longer any concern about  pulse timings or pulse-sequence engineering, and the higher driving frequencies that we can achieve with continuous fields are naturally expected to
eliminate higher frequency noise sources \cite{HanggiChemicalPhysics2004}. Indeed, continuous DD schemes to protect a single qubit have
attracted considerable interest \cite{HanggiChemicalPhysics2004, ChenPRA2006, FanchiniPRA12007,FanchiniPRA32007}.
Among the known features of single-qubit continuous DD, most relevant here
   is the fact that continuous control fields may be constructed to protect a quantum state and implement a gate at the same time, without the use of any overhead \cite{FanchiniPRA12007,FanchiniPRA32007} (thus forming a type of DCG~\cite{violaPRL2009,violaPRA2009}).
   However, it is imperative, considering the complexity of two-qubit decoherence, that such Hamiltonians be constructed for two-qubit gates as well.  After all, in the circuit model of quantum computation, two-qubit gates are of fundamental importance.

What is lacking currently is a completely general treatment of continuous DD for two-qubit systems. A recent study \cite{KurizkiPRA2011} considered the application of external fields to protect a multi-qubit system against a restricted class of dephasing and relaxation mechanisms. Some stimulating progress has also been made
in Ref.~\cite{FanchiniPRA22007,*FanchiniarXiv2010} .
However, therein only local noise terms were considered, which amounts to making a specific assumption applicable to only one class of system-environment coupling.  With such an assumption, continuous fields were constructed for the protection of two-qubit quantum states and two-qubit gates against decoherence \cite{FanchiniPRA22007}.  Nevertheless, as also seen below, if only local noise terms are considered, then the issue of decoherence control is somewhat quite analogous to single-qubit continuous DD and is not universal. The explicit task of this work is hence to extend the work in Ref.~\cite{FanchiniPRA22007} to cases with the most general system-environment coupling.   Our universal two-qubit continuous DD schemes presented below may be of great use if very high control fidelities are non-essential. Certainly, universal continuous DD schemes may be also combined with pulsed DD for hybrid DD schemes.

The organization of this paper is as follows. In Sec.\:II, we construct continuous driving fields that are able to protect an arbitrary two-qubit quantum state against its environment. Once we have such a scheme, we use this knowledge for the construction of two-qubit gates protected against the most general decoherence.  It is shown in Sec.\:III that
an explicitly constructed
control Hamiltonian can implement and protect two-qubit gates at the same time, and can therefore yield much better gate performance. In Sec.\:IV, in order to test these theoretical results, we introduce a non-Markovian master equation. By modeling the environment as one (or many) thermal bath(s) possessing an Ohmic spectral density, we show, in Sec.\:V, the results of our numerical simulations illustrating the excellent performance of our fields in protecting two-qubit states as well as in implementing two-qubit gates. Finally, Sec.\:VI concludes this paper.

\section{UNIVERSAL CONTINUOUS DYNAMICAL DECOUPLING}

We start off by considering the Hamiltonian of a two-qubit system interacting with its environment (modeled by - possibly more than one - thermal bath later),
\begin{equation}
H_{\mbox{\scriptsize tot}} = H_0 + H_{\text{B}} + H_{\mbox{\scriptsize SB}}
\end{equation}
where $H_0$ denotes the Hamiltonian of the two-qubit system, $H_{\text{B}}$ the Hamiltonian of the environment, and $H_{\mbox{\scriptsize SB}}$ is the interaction Hamiltonian between the system and its environment.

The most general form of $H_{\mbox{\scriptsize SB}}$ is given by \cite{MukhtarPRA12010}
\begin{equation}
H_{\mbox{\scriptsize SB}} = \sum_{k = 1}^3 B_k^{(1)} \sigma_k^{(1)} + \sum_{k = 1}^3 B_k^{(2)} \sigma_k^{(2)} + \sum_{k = 1, l =1}^3 B_{kl}^{(12)}\sigma_k^{(1)}\sigma_l^{(2)},
\label{mostnoise}
\end{equation}
where $\sigma_1 = \sigma_x, \sigma_2 = \sigma_y, \sigma_3 = \sigma_z$, and the $B$ operators denote arbitrary environment operators. Note that this form is considerably more complex than the local-environment interaction Hamiltonian considered in Ref. \cite{FanchiniPRA22007}, since nonlocal
terms such as $B_{kl}^{(12)}\sigma_{k}^{(1)}\sigma_{l}^{(2)}$ are now taken into account.

We now consider continuous driving fields applied to the system, whose effect is described by the Hamiltonian $H_c(t)$. Corresponding to $H_c(t)$, there is a unitary operator $U_c(t)$, given by the time-ordered exponential of $H_c(t)$ ($\hbar = 1$ and $\mathcal{T}$ is the time-ordering operator throughout),
\begin{equation}
U_c(t) = \mathcal{T} \exp \left[ -i \int_0^{t} H_c(s)\, ds \right].
\end{equation}
In order to achieve continuous DD, $U_c(t)$ must fulfill two criteria. The first is that it should be periodic in time with a period denoted by $t_c$, that is,
\begin{equation}
\label{periodiccondition}
U_c(t + t_c) = U_c(t).
\end{equation}
Secondly, in order to decouple the system from the environment, we hope to have
\begin{equation}
\label{decouplingcondition}
\int_{0}^{t_c} U_c^\dagger (t) H_{\mbox{\scriptsize SB}} U_c(t)\, dt = 0.
\end{equation}
Technically, these conditions can be derived using the Magnus expansion. For completeness, following the treatment given in Ref.~\cite{FacchiPRA2005}, we show that these conditions indeed lead to a low-order decoupling of the system from the environment. Ideas and notation introduced here will be used again when we explain the reasoning behind the construction of control fields for gate protection.

The Hamiltonian for the total system in the presence of the control fields can be written (in the `lab' frame) as
\begin{equation}
 H_{\mbox{\scriptsize tot}} = H_0 + H_c(t) + H_\text{B} + H_{\mbox{\scriptsize SB}} = H' + H_c(t),
 \end{equation}
where
\begin{equation}
 H' \equiv H_0 + H_\text{B} + H_{\mbox{\scriptsize SB}}.
 \end{equation}
Our goal is to see how a state evolves under the action of this total Hamiltonian, if conditions \eqref{periodiccondition} and \eqref{decouplingcondition} are satisfied. In order to do so, we transform to the frame of the control fields, that is, we rotate the basis by $U_c(t)$. Then, in this frame, a total system-environment state evolves under the action of the unitary time-evolution operator,
\begin{equation}
\tilde{U}_{\mbox{\scriptsize tot}} (t) = \mathcal{T} \exp \left[ -i \int_{0}^t \tilde{H}'(s)\, ds \right],
\end{equation}
where $\tilde{H}'(s) = U_c^\dagger(s)H'U_c(s)$.

At time $t = Nt_c$ ($N$ is a positive integer), because $\tilde{H}'(s)$ is periodic with period $t_c$, we have
\begin{equation}
\tilde{U}_{\mbox{\scriptsize tot}}(t) = \left[ \tilde{U}_{\mbox{\scriptsize tot}}(t_c)\right]^N,
\end{equation}
and
\begin{equation}
\tilde{U}_{\mbox{\scriptsize tot}}(t_c) = \mathcal{T} \exp \left[ -i \int_0^{t_c} \tilde{H}'(s)\, ds \right].
\end{equation}
The Magnus expansion \cite{Magnus1954} allows us to write (refer to Appendix A),
\begin{equation}
\tilde{U}_{\mbox{\scriptsize tot}}(t_c) = \exp \left[ -i t_c (\tilde{H}^{(0)} + \tilde{H}^{(1)} + \hdots ) \right],
\end{equation}
with,
\begin{equation}
\tilde{H}^{(0)} = \frac{1}{t_c} \int_0^{t_c} ds \; \tilde{H}'(s).
\end{equation}
We ignore the higher order terms since we are concerned with a low-order DD only. 

Now it is at this point that the condition Eq.\,\eqref{decouplingcondition} comes in. Because of this condition, i.e.,
$\int_{0}^{t_c} U_c^\dagger (t) H_{\mbox{\scriptsize SB}} U_c(t) dt = 0$, we can eliminate the $H_{\mbox{\scriptsize SB}}$ term in $\tilde{H}^{(0)}$. Therefore, $\tilde{H}^{(0)} = \bar{H} + H_B$, with
\begin{equation}
\bar{H} = \frac{1}{t_c} \int_0^{t_c} ds \; U_c^\dagger(s)H_0 U_c(s).
\end{equation}
We also note that $\bar{H}$ is independent of $t_c$. Since $U_c(t)$ is periodic in time with period $t_c$, we can write it as some function of $t/t_c$, say, $U_c'(t/t_c)$. Then,
\begin{eqnarray}
\bar{H} &=& \frac{1}{t_c} \int_0^{t_c} dt \; U_c'^\dagger(t/t_c)H_0 U_c'(t/t_c), \nonumber \\
& = & \int_0^1 dx \; U_c'^\dagger(x)H_0 U_c'(x),
\end{eqnarray}
with $x = t/t_c$. It follows that $\bar{H}$ is indeed not an explicit function of $t_c$.

Keeping in mind that $N = t/t_c$, we then find that to lowest 
order in $t_c$,
\begin{equation}
\tilde{U}_{\mbox{\scriptsize tot}}(t) \approx \left[ e^{-it_c \tilde{H}^{(0)}} \right]^{t/t_c} \approx e^{-i\bar{H}t} e^{-iH_Bt}.
\label{productform}
\end{equation}
Finally, transforming back to the original frame (the `lab' frame), we find that the unitary evolution operator in this frame is
\begin{equation}
U_{\mbox{\scriptsize tot}}(t) \approx U_c(t) e^{-i\bar{H}t} e^{-iH_Bt}.
\end{equation}
But, because of the condition $t = Nt_c$ and the periodicity of $U_c(t)$, $U_c(Nt_c)$ is just identity. Obviously, then, the system has been decoupled from the environment - they both evolve independently, since $\bar{H}$ acts only on the system Hilbert space, while $H_{\text{B}}$ acts only on the environment Hilbert space.
 Roughly speaking, we can understand this result by realizing that under the condition in Eq.~\eqref{decouplingcondition}, the system-environment interaction is averaged out in the frame of the control fields. 

\subsection{Suppression of local noise}

Let us now come back to our problem of finding control fields to protect an arbitrary two-qubit state. As stated before, in Ref.~\cite{FanchiniPRA22007}, continuous fields were found to eliminate the local noise terms.
Local noise here means the following restricted form of system-environment coupling,
\begin{eqnarray}
H_{\mbox{\scriptsize SB}} = \sum_{k = 1}^3 B_k^{(1)} \sigma_k^{(1)} + \sum_{k = 1}^3 B_k^{(2)} \sigma_k^{(2)}.
\end{eqnarray}
Other coupling terms in Eq.~(\ref{mostnoise}) different from above are loosely called nonlocal noise terms.

We recap what is already known - how to find continuous fields to eliminate the above-defined local noise terms, as is done in Ref.~\cite{FanchiniPRA22007} (but with more details). We first observe that with the unitary control operator
\begin{equation}
U_c(t) = U_c^{(1)}(t) U_c^{(2)}(t),
\end{equation}
where
\begin{equation}
U_c^{(k)}(t) = e^{-2\pi i \sigma_x^{(k)} n_x t/t_c}, \: k = 1,2,
\end{equation}
with $n_x$ a non-zero integer, we eliminate noise terms proportional to $\sigma_y^{(k)}$ and $\sigma_z^{(k)}$. Intuitively, this follows from the fact that $U_c(t)$ is just a rotation operator, and therefore causes the $\sigma_y^{(k)}$ and $\sigma_z^{(k)}$ noise terms to rotate so that they average out to zero. However, it leaves the $\sigma_x^{(k)}$ noise terms untouched. In order to cancel these noise terms as well, we modify our unitary control operator to
\begin{equation}
U_c^{(k)}(t) = e^{-2\pi i \sigma_x^{(k)} n_x t/t_c}e^{-2\pi i \sigma_z^{(k)} n_z t/t_c},
\end{equation}
where $n_z$ is another non-zero integer satisfying the condition $n_x \neq n_z$. The unitary control operator now consists of two rotation operators. The $\sigma_x^{(k)}$ part of the unitary operator rotates the $\sigma_z^{(k)}$ and $\sigma_y^{(k)}$ noise terms and averages them out to zero, while the $\sigma_z^{(k)}$ part of the unitary control operator takes care of the remaining $\sigma_x^{(k)}$ noise terms (it nevertheless also rotates the $\sigma_y^{(k)}$ operators). The condition $n_x \neq n_z$ is important because otherwise, the effect of the second rotation cancels some effect of the first rotation such that, for instance, the $\sigma_y^{(k)}$ noise terms do not average out to zero. All these claims can be examined by explicitly verifying if Eq.~\eqref{decouplingcondition} holds. For instance, we observe that
\begin{equation}
\int_0^{t_c} U_c^\dagger (t) \sigma_x^{(k)} U_c(t)\, dt = \int_0^{t_c} e^{4\pi i n_z \sigma_z^{(k)}t/t_c} \sigma_x^{(k)} dt = 0.
\end{equation}
Also,
\begin{eqnarray}
&&\int_0^{t_c} U_c^\dagger (t) \sigma_z^{(k)} U_c(t)\, dt \nonumber \\
& = &\int_0^{t_c} e^{2\pi i n_z \sigma_z^{(k)}t/t_c} e^{4\pi i n_x \sigma_x^{(k)}t/t_c} e^{-2\pi i n_z \sigma_z^{(k)}t/t_c} \sigma_z^{(k)} dt. \nonumber \\
\end{eqnarray}
Using the commutator
\begin{equation*}
[e^{2i \omega n_x \sigma_x^{(k)} t}, e^{-i \omega n_z \sigma_z^{(k)} t}] = -2 i \sin(2n_x\omega t)\sin(n_z\omega t) \sigma_y^{(k)} ,\end{equation*}
where $\omega \equiv \frac{2\pi}{t_c}$, we can simplify,
\begin{eqnarray}
&&\int_0^{t_c} U_c^\dagger (t) \sigma_z^{(k)} U_c(t)\, dt  \nonumber \\
&=& - \int_0^{t_c} 2i e^{2\pi i n_z \sigma_z^{(k)}t/t_c} \sin(2n_x \omega t) \sin(n_z\omega t) \sigma_y^{(k)} \sigma_z^{(k)} dt \nonumber\\
&&+ \ \int_0^{t_c} e^{4\pi i n_x \sigma_x^{(k)}t/t_c} \sigma_z^{(k)} dt,
\end{eqnarray}
which further simplifies to
\begin{eqnarray}
&&\int_0^{t_c} U_c^\dagger (t) \sigma_z^{(k)} U_c(t)\, dt  \nonumber \\
&=& -2i \int_0^{t_c} [\cos(n_z\omega t) + i\sin(n_z \omega t)\sigma_z^{(k)}]
\nonumber \\
&& \times\ [\sin(2n_x\omega t)\sin(n_z \omega t)] \sigma_y^{(k)} \sigma_z^{(k)} dt.
\end{eqnarray}
Now,
\begin{equation}
\int_0^{t_c} \sin^2(n_z \omega t) \sin(2n_x \omega t) dt = 0,
\end{equation}
no matter what the values of $n_x$ and $n_z$ are, but in order to have
\begin{equation}
\int_0^{t_c} \cos(n_z \omega t) \sin(2n_x \omega t) \sin(n_z \omega t) dt = 0,
\end{equation}
we require that $n_x \neq n_z$. Therefore, if $n_x \neq n_z$, the $\sigma_x^{(k)}$ noise terms are eliminated. Similarly, one can check that the $\sigma_y^{(k)}$ noise terms are also eliminated.

The necessary control field to implement the unitary control operator $U_{c}(t)$ can be found from the Schrodinger equation $i \frac{\partial{U_c}}{\partial{t}} = H_c(t)U_c(t)$. We then find
\begin{eqnarray}
\label{Hcpre}
H_c(t) &=& \sum_{i = 1}^2 \left\lbrace \omega n_x \sigma_x^{(i)} \right. \nonumber  \\
&+& \left. \omega n_z \left[\cos(2\omega n_x t) \sigma_z^{(i)} - \sin(2\omega n_x t) \sigma_y^{(i)}\right]\right\rbrace. \notag \\
\end{eqnarray}
One obvious aspect of this control field is that both qubits are addressed in exactly
 the same way.  The field configuration is also quite simple: it consists of a local static field and a local rotating field.

 However, the control Hamiltonian found above is not universal. In particular, it cannot eliminate all possible forms of system-environment coupling shown in Eq.~(\ref{mostnoise}).  For instance, consider the noise term proportional to $\sigma_x^{(1)}\sigma_x^{(2)}$. We find that
\begin{eqnarray}
&&\int_0^{t_c} U_c^\dagger (t) \sigma_x^{(1)} \sigma_x^{(2)} U_c(t)\, dt \nonumber  \\
& = &\int_0^{t_c} e^{4\pi i n_z \sigma_z^{(1)}t/t_c} \sigma_x^{(1)} e^{4\pi i n_z \sigma_z^{(2)}t/t_c} \sigma_x^{(2)} dt \nonumber  \\
& = &\int_0^{t_c} \left[ \cos(2n_z \omega t) + i\sin(2n_z \omega t)\sigma_z^{(1)}\right ]\nonumber \\
&& \times\ \left[\cos(2n_z \omega t) + i\sin(2n_z\omega t)\sigma_z^{(2)}\right]\sigma_x^{(1)}\sigma_x^{(2)} dt \neq 0. \nonumber \\
\end{eqnarray}
This is obvious because
$$ \int_0^{t_c} \cos^2(2n_z\omega t) \sigma_x^{(1)}\sigma_x^{(2)} dt \neq 0.$$
Therefore, the $\sigma_x^{(1)}\sigma_x^{(2)}$ noise term does not average out to zero if $\sigma_x^{(1)}$ and $\sigma_x^{(2)}$
are rotated at the same frequency.

\subsection{Universal protection of two-qubit states}

We have just shown that it is not possible to eliminate all the noise terms by applying the same field to both qubits. So the important question is the following: is it possible to find a field configuration in which, by applying different fields to the two qubits, all the noise terms as shown in Eq.~(\ref{mostnoise}) can be eliminated?  At the same time, we would also like to retain the relative simplicity of the field configuration used for previous
local noise considerations. This motivates us to investigate if
\begin{equation}
\label{unitarycontroloperator}
U_c(t) = U_c^{(1)}(t) U_c^{(2)}(t),
\end{equation}
where
\begin{equation}
U_c^{(k)}(t) = e^{-2\pi i \sigma_x^{(k)} n_x^{(k)} t/t_c} e^{-2\pi i \sigma_z^{(k)} n_z^{(k)} t/t_c},
\end{equation}
serves to eliminate all the noise terms. Note that, since we allow the possibility of different fields being applied to the two qubits, this $U_c(t)$ differs from the previous $U_c(t)$ in that, this time, the integers in $U_c^{(1)}$ and $U_c^{(2)}$ need not be the same (previously we had $n_x^{(1)} = n_x^{(2)}$ and $n_z^{(1)} = n_z^{(2)}$). The postulated $U_c(t)$ is obviously periodic in time with period $t_c$. Furthermore, as shown below, we find that all the noise terms can indeed be eliminated, provided that the integers $n_x^{(1)}$, $n_z^{(1)}$, $n_x^{(2)}$, and $n_z^{(2)}$ fulfill some criteria. For simplicity, we consider $n_x^{(1)}$, $n_z^{(1)}$, $n_x^{(2)}$, and $n_z^{(2)}$ to be positive integers. Also, since we expect that the integers are different, for our own convenience in narrowing down the criteria fulfilled by the integers, we impose the condition
\begin{equation}
\label{orderingofintegers}
n_x^{(1)} < n_z^{(1)} < n_x^{(2)} < n_z^{(2)}.
\end{equation}

Let us now find the criteria that the integers $n_x^{(1)}$, $n_z^{(1)}$, $n_x^{(2)}$, and $n_z^{(2)}$ need to fulfill. In order to do this rigorously, we need to check that each noise term averages out to zero under the action of the applied fields. Since we have ordered the integers as in Eq.~\eqref{orderingofintegers}, we already have that,
\begin{equation}
n_x^{(1)} \neq n_z^{(1)},
\end{equation}
\begin{equation}
n_x^{(2)} \neq n_z^{(2)}.
\end{equation}
Using the derivations in Sec. II-A, it is easy to see that all local noise terms are indeed eliminated.

We next examine the fate of nonlocal noise terms. For instance, let us consider the noise term proportional to $\sigma_x^{(1)} \sigma_x^{(2)}$. This time we have,
\begin{eqnarray}
& & \int_0^{t_c} U_c^\dagger (t) \sigma_x^{(1)} \sigma_x^{(2)} U_c(t)\, dt \nonumber \\
& = &  \int_0^{t_c} e^{4\pi i n_z^{(1)} \sigma_z^{(1)}t/t_c} \sigma_x^{(1)} e^{4\pi i n_z^{(2)} \sigma_z^{(2)}t/t_c} \sigma_x^{(2)} dt \nonumber  \\
& = & \int_0^{t_c} \left[\cos(2n_z^{(1)}\omega t) + i\sin(2n_z^{(1)}\omega t)\sigma_z^{(1)}\right] \nonumber \\
&& \times\  \left[\cos(2n_z^{(2)}\omega t) + i\sin(2n_z^{(2)}\omega t)\sigma_z^{(2)}\right]\sigma_x^{(1)}\sigma_x^{(2)} dt, \nonumber \\
\end{eqnarray}
which is zero, provided that,
\begin{equation}
n_z^{(1)} \neq n_z^{(2)}.
\end{equation}
This is obvious because under the condition $n_z^{(1)} \neq n_z^{(2)}$,
$$ \int_0^{t_c} \cos(2n_z^{(1)}\omega t)\cos(2n_z^{(2)}\omega t)\,dt$$ and similar terms
are all zero.

Therefore, one observes that if $\sigma_x^{(1)}$ and $\sigma_x^{(2)}$ are rotated at different frequencies, then the noise term proportional to $\sigma_x^{(1)}\sigma_x^{(2)}$ is eliminated. This condition gives support to our intuition that the fields applied to each qubit should be different.

We next outline the calculation for the noise term proportional to $\sigma_z^{(1)} \sigma_z^{(2)}$. In this case, the calculation is considerably more involved. To calculate the required integral involving this noise term, we first calculate (suppressing the $k$ index),
\begin{eqnarray}
& & e^{2\pi i n_z \sigma_z t/t_c} e^{2\pi i n_x \sigma_x t/t_c} \sigma_z e^{-2\pi i n_x \sigma_x t/t_c} e^{-2\pi i n_z \sigma_z t/t_c} \nonumber \\
& = & e^{2\pi i n_z \sigma_z t/t_c} e^{4\pi i n_x \sigma_x t/t_c} e^{-2\pi i n_z \sigma_z t/t_c}  \sigma_z \nonumber \\
& = & \cos(2n_x\omega t) \sigma_z + \sin(2n_x\omega t) \sin(2n_z\omega t) \sigma_x \nonumber \\
&& +\ \sin(2n_x\omega t)\cos(2n_z\omega t) \sigma_y .
\end{eqnarray}
The integral that we wish to set to zero then becomes
\begin{eqnarray}
\label{nonlocalnoisecancelexample}
&& \int_0^{t_c} U_c^\dagger (t) \sigma_z^{(1)} \sigma_z^{(2)} U_c(t)\, dt  \nonumber \\
&=&\int_0^{t_c} \left[\cos(2n_x^{(1)}\omega t) \sigma_z^{(1)} + \sin(2n_x^{(1)}\omega t) \sin(2n_z^{(1)}\omega t) \sigma_x^{(1)}\right. \nonumber \\
&& +\ \left.\sin(2n_x^{(1)}\omega t) \cos(2n_z^{(1)}\omega t) \sigma_y^{(1)}\right] \nonumber \\
& &\times\ \left[\cos(2n_x^{(2)}\omega t) \sigma_z^{(2)}  +  \sin(2n_x^{(2)}\omega t) \sin(2n_z^{(2)}\omega t) \sigma_x^{(2)}\right. \nonumber \\
&& +\ \left.\sin(2n_x^{(2)}\omega t)
\cos(2n_z^{(2)}\omega t) \sigma_y^{(2)}\right]\   dt.
\end{eqnarray}
By multiplying the terms in the square brackets above, we get different terms. Each of these terms must individually integrate to zero, because the tensor products of two Pauli matrices are linearly independent in the operator space. So, for example, we require that
\begin{equation}
\int_0^{t_c} \cos(2n_x^{(1)}\omega t) \sin(2n_x^{(2)}\omega t) \sin(2n_z^{(2)}\omega t) \sigma_z^{(1)}\sigma_x^{(2)} dt = 0,
\end{equation}
which is true provided that
\begin{equation}
 n_x^{(1)} + n_x^{(2)} - n_z^{(2)} \neq 0.
 \end{equation}
One might think that we would also need three other conditions, one of which is given by
\begin{equation} n_x^{(1)} - n_x^{(2)} + n_z^{(2)} \neq 0.
\end{equation}
Fortunately,  due to the ordering to the integers in Eq.~\eqref{orderingofintegers}, this condition and the other two
are redundant. Therefore, we can ignore these redundant conditions.

Similarly, analyzing each of the others terms in Eq.\,\eqref{nonlocalnoisecancelexample} one by one, and keeping the ordering of the integers in mind, we arrive at the following list of criteria:
\begin{eqnarray}
n_z^{(2)} \neq n_x^{(1)} + n_x^{(2)}, \notag \\
n_x^{(2)} \neq n_z^{(1)} + n_x^{(1)}, \notag \\
n_x^{(1)} + n_z^{(1)} + n_x^{(2)} - n_z^{(2)} \neq 0,\notag \\
n_x^{(1)} - n_z^{(1)} - n_x^{(2)} + n_z^{(2)} \neq 0.
\end{eqnarray}

The other seven types of system-environment coupling shown in Eq.~(\ref{mostnoise}) can be treated in a similar fashion and will not be repeated here.
Carefully going through all of them, we come to the conclusion that by applying the unitary control operator in Eq.\,\eqref{unitarycontroloperator} with the following conditions,
\begin{align}
n_x^{(1)} < n_z^{(1)} < n_x^{(2)} < n_z^{(2)}, \notag \\
n_x^{(2)} \neq n_x^{(1)} + n_z^{(1)}, \notag \\
n_z^{(2)} \neq n_x^{(1)} + n_x^{(2)}, \notag \\
n_z^{(2)} \neq n_x^{(1)} + n_z^{(1)}, \notag \\
n_z^{(2)} \neq n_z^{(1)} + n_x^{(2)}, \notag \\
n_x^{(1)} + n_z^{(1)} + n_x^{(2)} - n_z^{(2)} \neq 0, \notag \\
n_x^{(1)} - n_z^{(1)} - n_x^{(2)} + n_z^{(2)} \neq 0,
\end{align}
our two-qubit system can be (approximately) decoupled from the most general environment. From the above conditions,
it is seen that not only must the frequencies in $U_c(t)$ be all different, but also that neither of the two larger frequencies should be the sum of two smaller frequencies. Furthermore, the difference of the two larger frequencies should not be equal to the sum or the difference of the two smaller frequencies.
It is not hard to find integers that fulfill all the criteria we have found. One possible choice is $ n_x^{(1)} = 1, n_z^{(1)} = 2,  n_x^{(2)} = 4, n_z^{(2)} = 8$.

Finally, the control Hamiltonian, $H_c(t)$, which is needed to generate the unitary operator $U_c(t)$, is found to be [from the time derivative of $U_c(t)$]
\begin{eqnarray}
\label{Hc}
H_c(t) &=& \sum_{i = 1}^2 \left\{ \omega n_x^{(i)} \sigma_x^{(i)}\right. \notag \\
&+& \left.\omega n_z^{(i)} \left[\cos(2\omega n_x^{(i)} t) \sigma_z^{(i)} - \sin(2\omega n_x^{(i)} t) \sigma_y^{(i)}\right]\right\}. \notag \\
\end{eqnarray}
Each of the two qubits is now subject to a different local control field consisting of a static field and a rotating field - we must address each qubit individually. With these control fields, the two-qubit system is dynamically decoupled from the environment, for all possible types of system-environment coupling.
Note also from Eq.~(\ref{Hc}) that the field amplitude should also go up if the frequencies of the driving field are increased to
compete with the cutoff frequency of the environment.

\section{PROTECTION OF TWO-QUBIT GATES}

Once we have the control operator $U_c(t)$ that is able to protect two qubits against decoherence in a universal manner,
the next natural question is how to turn on coherent evolution in two-qubit systems such that two-qubit gates can be also
protected.
This is important because, in reality, there is no instantaneous quantum gate. As shown below, we can extend
our previous considerations to protect a two-qubit state and implement a desired gate at the same time.  Some early studies considered
the protection of a two-qubit gate against random dephasing \cite{KurizkiPRA2007} and against bit-flip errors \cite{HanggiPRL2005}, but
these early decoherence suppression approaches are not applicable to an arbitrary environment.
Our procedure is analogous to Ref.~\cite{FanchiniPRA12007,FanchiniPRA32007}, but for most general system-environment coupling in two-qubit systems.
The extension here is worthwhile because in actual realizations of two-qubit gates, it is unavoidable
that the two-qubit interaction Hamiltonian will suffer from fluctuations, on top of local noise terms seen by each individual qubit.

\subsection{Two-qubit gate under pure dephasing}
To illustrate the method, we start off with the simple case of pure dephasing. The interaction between the two qubits and their environment
is given by, \begin{equation}
H_{\mbox{\scriptsize dephasing}} = B_z^{(1)} \sigma_z^{(1)} + B_z^{(2)} \sigma_z^{(2)} + B_{zz}^{(12)} \sigma_z^{(1)} \sigma_z^{(2)} .
\end{equation}
As can be easily verified, in this case, a simpler control operator
\begin{equation}
U_c(t) = \exp(-2\pi i \sigma_x^{(1)} n_1 t/t_c) \exp(-2\pi i \sigma_x^{(2)} n_2 t/t_c),
\end{equation}
with $ n_1 \neq n_2 $ suffices to protect two-qubit states.

Consider now a two-qubit gate that converts a separable state into a Bell state, i.e.,
$$ \ket{\psi_0} = \frac{1}{\sqrt{2}} ( \ket{\uparrow}_x + \ket{\downarrow}_x ) \ket{\downarrow}_x \longrightarrow \frac{1}{\sqrt{2}} ( \ket{\uparrow \downarrow}_x + \ket{\downarrow \uparrow}_x ), $$
where
\begin{eqnarray}
\ket{\uparrow}_x = \frac{1}{\sqrt{2}}(\ket{0} + \ket{1}), \\
\ket{\downarrow}_x = \frac{1}{\sqrt{2}}(\ket{0} - \ket{1}),
\end{eqnarray}
with $\ket{0}$ and $\ket{1}$ being eigenstates of the $\sigma_z$ operator. We consider the initial state to be $\ket{\psi_0}$ in order to bring out the effect of the dephasing noise clearly. It should be noted that the above gate is analogous to the usual controlled-NOT (CNOT) gate, since the CNOT gate performs the operation $\frac{1}{\sqrt{2}} ( \ket{0} + \ket{1} ) \ket{1} \longrightarrow \frac{1}{\sqrt{2}} ( \ket{01} + \ket{10} )$. Therefore, we refer to the gate implementing the above operation as the $\overline{\mbox{CNOT}}$ gate. We work with the $\overline{\mbox{CNOT}}$ gate because it generates entanglement - the usual CNOT gate acting on $\frac{1}{\sqrt{2}} ( \ket{\uparrow}_x + \ket{\downarrow}_x ) \ket{\downarrow}_x$ yields a separable state.

The most straightforward way to implement the $\overline{\mbox{CNOT}}$ gate (up to an irrelevant global phase) is to use the two-qubit Hamiltonian,
\begin{equation}
\label{CNOTsimplefield}
H_0 = \frac{\pi}{2\tau} \frac{1}{2}\left( \sigma_x^{(1)} + \sigma_z^{(2)} - \sigma_x^{(1)}\sigma_z^{(2)}\right),
\end{equation}
where $\tau$ is the time over which the gate is implemented. Note that no decoherence control fields are being applied at this stage. Therefore, during the gate operation time, the two-qubit state is vulnerable to decoherence due to the environment. Our task is to modify the Hamiltonian given by Eq.~\eqref{CNOTsimplefield} such that the new Hamiltonian not only implements the $\overline{\mbox{CNOT}}$ gate, but also prevents decoherence. 

In order to find this new Hamiltonian, we begin by writing the system Hamiltonian as
\begin{equation}
H_{\text{S}}(t) = H_0(t) + H_c(t),
\end{equation}
where $H_0(t)$ implements the gate.  In order to find the unitary control operator that both implements the gate and protects against decoherence, the basic idea is to once again transform to the frame given by $H_c(t)$.  Now, in this frame, the effect of the environment has already been largely removed - it is almost as if the environment were not there. Therefore, we implement the gate in this picture. After doing so, we simply transform back to our original reference frame to find the total unitary control operator.

Let us now carry out these ideas in detail in order to find the required $H_{\text{S}}(t)$. We first, once again, write the total Hamiltonian as
\begin{equation}
H = H'(t) + H_c(t),
\end{equation}
where $H'(t) = H_0(t) + H_{\text{B}} + H_{\mbox{\scriptsize SB}}$. We now transform to the frame of the control fields, as we did before. In this frame, $H_0(t)$ becomes $\tilde{H}_0 = U_c^\dagger(t)H_0(t)U_c(t)$. Corresponding to this Hamiltonian, there is a unitary time-evolution operator,
\begin{equation}
\tilde{U}_0(t) = \mathcal{T} \exp \left[ -i \int_0^t \tilde{H}_0(s) ds \right].
\label{u0t}
\end{equation}
It is this unitary operator that we use to implement the gate. Therefore,
\begin{equation}
\label{CNOTtildefreeunitaryoperator}
\tilde{U}_0(t) = \exp\left[ -i\frac{\pi}{2\tau}\frac{t}{2}(I + \sigma_x^{(1)} + \sigma_z^{(2)} - \sigma_x^{(1)}\sigma_z^{(2)})\right],
\end{equation}
where again $\tau$ is the time over which the gate is implemented up to a global phase, that is, $U_{\mbox{\scriptsize gate}} = \tilde{U}_0(t = \tau)$. We set $\tau = Nt_c$ ($N$ is a positive integer).  Comparing Eqs.~(\ref{u0t}) and (\ref{CNOTtildefreeunitaryoperator}), it is clear that
\begin{equation}
\tilde{H}_0 = \frac{\pi}{2\tau} \frac{1}{2}\left( \sigma_x^{(1)} + \sigma_z^{(2)} - \sigma_x^{(1)}\sigma_z^{(2)}\right),
\end{equation}
will do the right job (this choice for  $\tilde{H}_0$ is simple because it is time-independent).

As shown in our previous general consideration of DD in Sec.~II,
the total system-environment time-evolution operator in the frame of the control fields is already approximately
 decoupled into a product of system and environment parts [see Eq.~(\ref{productform})]. In particular, applying the Magnus
 expansion to the following total evolution operator
\begin{equation}
\tilde{U}_{\mbox{\scriptsize tot}}(\tau) = \mathcal{T} \exp \left[ -i \int_0^\tau \tilde{H}'(s)\, ds \right],
\end{equation}
we have that for sufficiently small $t_c$, 
\begin{equation}
\tilde{U}_{\mbox{\scriptsize tot}}(\tau) \approx e^{-i\tilde{H}_0 \tau} e^{-iH_{\text{B}} \tau} = U_{\mbox{\scriptsize gate}} e^{-iH_{\text{B}}\tau}.
\end{equation}
We finally transform back to the original frame. In this frame, the unitary time evolution operator is given by
\begin{equation}
U_{\mbox{\scriptsize tot}}(\tau) \approx  U_c(\tau) U_{\mbox{\scriptsize gate}} e^{-iH_{\text{B}}\tau}.
\end{equation}
But $U_c(\tau)$ is just identity, leading to
\begin{equation}
U_{\mbox{\scriptsize tot}}(\tau) \approx U_{\mbox{\scriptsize gate}} e^{-iH_{\text{B}}\tau}.
\end{equation}
Clearly then, the desired gate operation is performed on the two-qubit system.

For arbitrary time $t$, the system is also approximately decoupled from the environment
(because in the limit $t_c\approx 0$, $t$ is always close to an integer multiple of $t_c$).
Then, in the lab frame the overall unitary evolution operator for the two-qubit system at arbitrary time $t$ is given by
\begin{eqnarray}
U_{\text{S}}(t) & = & U_c(t) \tilde{U}_0(t)\nonumber \\
&=& \exp(-2\pi i \sigma_x^{(1)} n_1 t/t_c) \exp(-2\pi i \sigma_x^{(2)} n_2 t/t_c) \tilde{U}_0(t). \nonumber \\
\end{eqnarray}
Further using the time-dependent Schrodinger equation, the Hamiltonian that generates the overall evolution operator $U_{\text{S}}(t)$ can be obtained as follows,
\begin{eqnarray}
\label{CNOTHamiltonian}
 H_{\text{S}}(t) &=& \omega n_1 \sigma_x^{(1)} + \omega n_2 \sigma_x^{(2)} + \frac{\pi}{2 \tau} \frac{1}{2} \left[ \sigma_x^{(1)} \right. \notag \\
 &+& \left. \sigma_z^{(2)} \cos (2\omega n_2 t) - \sigma_y^{(2)} \sin (2 \omega n_2 t)  \right. \notag \\
 &-& \left. \sigma_x^{(1)} \sigma_z^{(2)} \cos(2\omega n_2 t) + \sigma_x^{(1)} \sigma_y^{(2)} \sin (2 \omega n_2 t)\right], \notag \\
\end{eqnarray}
with $ n_1 \neq n_2 $.  By our construction above, such a
field configuration implements the gate and protects against two-qubit pure-dephasing  at the same time.  Note that
here some nonlocal field components are needed. This is expected. After all, the original $\overline{\mbox{CNOT}}$ gate Hamiltonian, Eq.~\eqref{CNOTsimplefield}, also needs a qubit-qubit interaction term. The message is that an oscillating qubit-qubit interaction
can be highly useful in implementing robust two-qubit gates in a noisy environment.  This need for oscillating qubit-qubit interaction
here should not be regarded as a great disadvantage of our universal continuous DD. In fact,
  even in pulsed DD schemes for entanglement protection \cite{MukhtarPRA12010},
  pulsed qubit-qubit interaction is necessary to reduce the number of UDD layers.  The requirement for time-dependent qubit-qubit
  interaction terms is also consistent with previous case studies under the general DCG framework \cite{violaPRL2009,violaPRA2009}.


\subsection{CZ gate protected against most general environment}
We are now ready to carry out similar calculations to  construct a field configuration that protects a two-qubit gate against all possible forms of
system-environment coupling.  As an example, we consider the implementation of a  controlled phase (CZ) gate \cite{Brussbook}. This case is representative because,
if we can reliably implement the CZ gate in the presence of arbitrary decoherence sources, then, together with single-qubit gates,
we can perform universal gate operations in the presence of an unknown environment.

The CZ gate, in a matrix form in the standard representation, can be written as,
\begin{equation}
U_{\text{CZ}} = \left( \begin{array}{cccc} 1 & 0 & 0 & 0  \\ 0 & 1 & 0 & 0 \\ 0 & 0 & 1 & 0 \\ 0 & 0 & 0 & -1 \end{array}  \right).
\end{equation}
In order to achieve this unitary operation (up to a global phase) on a two-qubit system, a desired bare Hamiltonian would be,
\begin{equation}
\label{CZsimplefield}
H_0 = \frac{\pi}{2\tau} \frac{1}{2}\left( \sigma_z^{(1)} + \sigma_z^{(2)} - \sigma_z^{(1)}\sigma_z^{(2)}\right).
\end{equation}
where $\tau$ is the gate operation time. But the resulting two-qubit state is not protected against the environment.
As such, we seek instead a time-dependent system Hamiltonian $H_{\text{S}}(t)$.

Our previous treatment for the $\overline{\mbox{CNOT}}$ gate in a pure dephasing model can be extended easily. The physical picture underlying
the technique remains exactly the same.  That is, we implement the desired gate in the rotating frame and then transform it back to
the lab frame.
Following our previous notation, in the rotating frame we hope to have,
\begin{equation}
\label{CZgatefreeunitaryoperator}
\tilde{U}_0(t) = \exp\left[ -i\frac{\pi}{2\tau}\frac{t}{2}(I + \sigma_z^{(1)} + \sigma_z^{(2)} - \sigma_z^{(1)}\sigma_z^{(2)})\right].
\end{equation}
On the other hand, $U_c(t)$ is given by Eq.\,\eqref{unitarycontroloperator}. By combining these two unitary operators as before, the sought Hamiltonian is determined by simply using the time-dependent Schrodinger equation on the overall unitary evolution operator. It is
found to be,
\begin{eqnarray}
\label{CZgateHamiltonian}
H_{\text{S}}(t) &=& \sum_{i = 1}^2 \left\lbrace \omega n_x^{(i)} \sigma_x^{(i)} + \omega n_z^{(i)} \left[\cos(2\omega n_x^{(i)} t) \sigma_z^{(i)} \right. \notag \right. \\ \notag
&-& \left. \left. \sin(2\omega n_x^{(i)} t)\sigma_y^{(i)}\right]\right\rbrace + \frac{\pi}{2\tau} \frac{1}{2} \left[ \sigma_z^{(1)} \cos(2\omega n_x^{(1)} t) \right. \\ \notag
&-& \left. \sigma_y^{(1)} \sin(2\omega n_x^{(1)} t) + \sigma_z^{(2)} \cos(2\omega n_x^{(2)} t) \right. \\ \notag
&-& \left. \sigma_y^{(2)} \sin(2\omega n_x^{(2)} t)  \right. \\ \notag
&-& \left. \sigma_z^{(1)} \sigma_z^{(2)} \cos(2\omega n_x^{(1)} t) \cos(2\omega n_x^{(2)} t) \right. \\ \notag
&+& \left. \sigma_z^{(1)} \sigma_y^{(2)} \cos(2\omega n_x^{(1)} t) \sin(2\omega n_x^{(2)} t)  \right.  \\ \notag
&+& \left. \sigma_y^{(1)} \sigma_z^{(2)} \sin(2\omega n_x^{(1)} t) \cos(2\omega n_x^{(2)} t) \right. \\
&-& \left. \sigma_y^{(1)} \sigma_y^{(2)} \sin(2\omega n_x^{(1)} t) \sin(2\omega n_x^{(2)} t)\right].
\end{eqnarray}
We stress that here we did not make any assumption about the system-environment coupling.
A CZ gate can hence be implemented and protected against any type of decoherence, so long as the driving frequencies are sufficiently large (also sufficiently strong)
relative to the cutoff frequency of the environment.
 Comparing the system Hamiltonian here with that in the previous pure-dephasing case,  more oscillating qubit-qubit
 interaction terms are required for decoherence suppression.  Another interesting observation is that here, the oscillating qubit-qubit interaction
 terms carry the sum and the difference frequencies $2(n_x^{(1)}+ n_x^{(2)})\omega$ and $2(n_x^{(1)}- n_x^{(2)})\omega$. This feature can be regarded as
a result of the dual role of the control fields (implementing and protecting a gate). It
 is also consistent with the fact that the two qubits should be rotated at different frequencies.  We do not suggest that the required control fields
 in Eq.~(\ref{CZgateHamiltonian}) are easy to realize experimentally. But at least, such an explicit solution as an example of universal DD
 is indicative of what could be
 crucial in protecting two-qubit gates without making assumptions of system-environment coupling.
 Our two-qubit gate construction also constitutes an explicit and simple implementation of DCG \cite{violaPRL2009,violaPRA2009} to fight against \emph{arbitrary} (environment-induced) single-qubit and two-qubit errors, using a static field plus several continuous-wave driving fields of different frequencies.


\section{THE MASTER EQUATION}

In this section, we briefly introduce a master equation for our use in numerical simulations. Here we only summarize the basic formalism. For a conceptually simple derivation of the master equation based solely on time-dependent perturbation theory and not on any special techniques such as projection-operator methods, please see Appendix B.

In the master equation approach \cite{BPbook}, we find the dynamics of a system that is coupled to the environment by  first considering the total system consisting of the system and the environment as closed. We then trace over the environment to obtain a differential equation, known as the master equation, for the reduced
density matrix of the system.

We start by writing down the Hamiltonian of the total system as
\begin{equation}
H_{\mbox{\scriptsize tot}} = H_{\text{S}}(t) + H_{\text{B}} + H_{\mbox{\scriptsize SB}}(t),
\end{equation}
where $H_\text{S}(t)$ describes the system Hamiltonian, $H_\text{B}$ the environment Hamiltonian, and $H_{\mbox{\scriptsize SB}}$ is the interaction between the system and the environment. In general, we can write
\begin{equation}
H_{\mbox{\scriptsize SB}} (t) = \sum_j F_j(t) \otimes B_j(t).
\end{equation}
Here, the $F_j(t)$ are operators in the system Hilbert space, and the $B_j(t)$ are operators in the environment Hilbert space.

We assume that the interaction between the system and the environment is weak, and that the total initial state of the system and environment is a product state. We can then derive the master equation describing the time evolution of the reduced density matrix $\rho$ of the system as,
\begin{align}
\label{masterequation}
\frac{d\rho(t)}{dt} &= i [\rho(t),H(t)] + \notag \\
&\sum_j \int_{t_0}^t ds \lbrace [\bar{F}_j(t,s)\rho(t),F_j(t)]C_{ts}^j + h.c. \rbrace,
\end{align}
where,
\begin{equation}
\bar{F}_j(t,s) = U_{\text{S}}(t,s)F_j(s)U_{\text{S}}^\dagger(t,s),
\end{equation}
\begin{equation}
C_{ts}^j = \langle \tilde{B}_j(t) \tilde{B}_j (s) \rangle,
\end{equation}
\begin{equation}
\tilde{B}_j(t) = U_{\text{B}}^\dagger(t,t_0)  B(t) U_{\text{B}}(t,t_0),
\end{equation}
and $U_{\text{B}}(t,t_0)$ and $U_{\text{S}}(t,t_0)$ are the unitary time-evolution operators corresponding to $H_{\text{B}}$ and $H_{\text{S}}(t)$ respectively. Such a master equation has been used previously in Ref. \cite{KurizkiPRL2004}.

We consider the environment as a collection of an infinite number of harmonic oscillators, so that
\begin{equation}
H_\text{B} = \sum_j \sum_k \omega_{j,k} a^{\dagger}_{j,k} a_{j,k}.
\end{equation}
Here index $k$ denotes different modes of the oscillators in one bath, and
index $j$ denotes different thermal baths. Furthermore, we take the $B_j$ operators for the $j$th bath as
\begin{equation}
B_j = \sum_k (g_{j,k} a_{j,k} + g_{j,k}^{*} a_{j,k}^\dagger),
\end{equation}
where the $g_{j,k}$ are coupling strength parameters.  All the baths are assumed to be in a thermal equilibrium state with the same temperature $T$.
We then note that the bath correlation function, given by $C_{ts}^j$, can be written as a function of the time difference $t - s$. In order to proceed with the calculation the bath correlation functions, the discrete modes of the environment are replaced by a smooth continuum of modes specified by the so-called
spectral density $J(\Omega)$. For our numerical simulations, we consider an Ohmic spectral density with an exponential cutoff, that is,
\begin{equation}
J(\Omega) = G \Omega e^{-\Omega/\omega_c},
\end{equation}
where $G$ is the coupling constant, and $\omega_c$ is the cutoff frequency of a bath. For simplicity, we assume that all baths have the same spectral density, with the same $G$ and $\omega_c$.

\section{NUMERICAL RESULTS}

In this section we present numerical results illustrating universal protection of two-qubit states and two-qubit gates.
Two extreme cases are considered.  In the first case, two qubits are coupled to fifteen different baths, each of which induces one type of
 system-environment coupling. The operators $F_j$ in the master equation \eqref{masterequation} are then given by,
\begin{equation}
\label{differentbathcoupling}
F_j = \frac{1}{2} \sigma_k^{(1)}\sigma_l^{(2)},
\end{equation}
where $\sigma_0 = I$,  and the notation for Pauli matrices is the same as before.
In the second case,  all possible types of decoherence are modeled by a common bath. That is, in the master equation [see Eq.~\eqref{masterequation}],
there is only one $F_j$, which is written as,
\begin{equation}
\label{samebathcoupling}
F = \frac{1}{2}\left(\sum_{k = 1}^3 \sigma_k^{(1)} + \sum_{k = 1}^3 \sigma_k^{(2)} + \sum_{k = 1, l =1}^3 \sigma_k^{(1)}\sigma_l^{(2)}\right).
\end{equation}
In addition, pure-dephasing cases are also considered, with the different and common bath cases defined in an analogous way. The only difference is that for
pure-dephasing cases, there are no terms containing $\sigma_x$ or $\sigma_y$ in the $F_j$ operators.

We work in dimensionless units with $\hbar = 1$ and $k_BT = 2$. In these units, the parameters we use are $\omega_C = 2\pi$, and $t_c = 0.5$ (so $\omega = 4\pi$), unless stated otherwise. As the measure of bipartite entanglement, we use the concurrence \cite{WoottersPRL1998}. Given a two-qubit density matrix $\rho$, the concurrence, $C$, is defined as
$ C \equiv \max\lbrace \lambda_1 - \lambda_2 - \lambda_3 - \lambda_4, 0 \rbrace$,
where the $\lambda_i$ are the square roots of the eigenvalues (in descending order) of the matrix
$ \rho (\sigma_y^{(1)} \otimes \sigma_y^{(2)})\rho^{*}(\sigma_y^{(1)} \otimes \sigma_y^{(2)})$ (the asterisk denotes complex conjugation).

We first present results of two-qubit state protection using our universal continuous DD fields.
For convenience, the self-Hamiltonian of the two-qubit system is set to zero.
\begin{figure}[H]
   \includegraphics{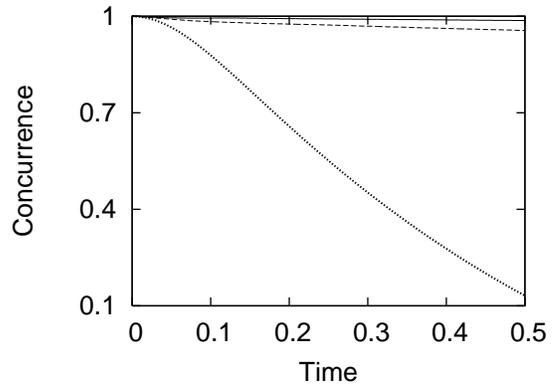}
   \centering
  	\caption{Entanglement vs time (in dimensionless units) without decoherence control fields (dotted line) and with applied control
   fields (dashed and solid lines). The environment is modeled by 15 different baths, i.e.,  in our master equation, the system coupling operators are
   given by Eq.\,\eqref{differentbathcoupling}. The dashed line is for $n_x^{(1)} = 1, n_z^{(1)} = 2,  n_x^{(2)} = 4, n_z^{(2)} = 8$.  The solid line is for stronger and higher frequency fields with $n_x^{(1)} = 2, n_z^{(1)} = 4,  n_x^{(2)} = 8, n_z^{(2)} = 16$.  For this numerical simulation, we use $G = 0.05$.}
  	\label{differentbathprotection}
\end{figure}

\begin{figure}[H]
   \includegraphics{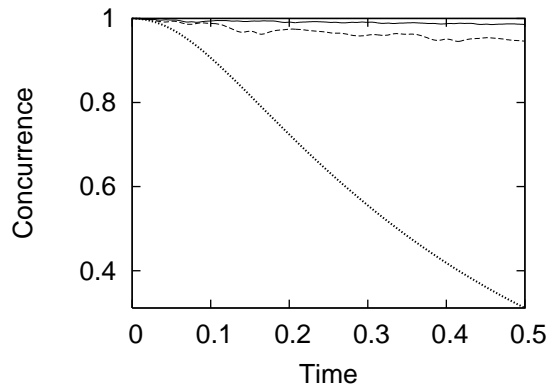}
   \centering
  	\caption{Entanglement versus time (in dimensionless units) without decoherence control fields (dotted line) and with applied control fields (dashed and solid lines). Here the environment is modeled by one common bath, with the system coupling operator given by Eq.\,\eqref{samebathcoupling}. The parameters used are the same as in Fig.~\ref{differentbathprotection}.}
  	\label{samebathprotection}
\end{figure}

Figures \ref{differentbathprotection} and \ref{samebathprotection} illustrate the performance of the decoherence control fields
in protecting two-qubit entanglement against an environment that generates all types of decoherence.
  Without these fields, we see (dotted curve) that, in both the common-bath and different-bath cases,
the entanglement rapidly decays due to the interaction with the environment. The situation changes dramatically after switching on the continuous
control fields. The dashed curves demonstrate the suppression of entanglement decay due to the control fields. Furthermore, by applying fields of greater strength and higher frequency (the solid curves) - thus effectively reducing $t_c$ - even better protection of entanglement is achieved.

We now study the effectiveness of the control Hamiltonian found in  Eq.\,\eqref{CNOTHamiltonian} in implementing the $\overline{\mbox{CNOT}}$ gate in the presence of pure dephasing.  First of all, the Hamiltonian that only implements the gate without decoherence control is given by Eq.\,\eqref{CNOTsimplefield}, which we rewrite here for convenience,
$$H_0 = \frac{\pi}{2\tau} \frac{1}{2}\left( \sigma_x^{(1)} + \sigma_z^{(2)} - \sigma_x^{(1)}\sigma_z^{(2)}\right).$$
This Hamiltonian should be contrasted with the Hamiltonian in Eq.\,\eqref{CNOTHamiltonian} that both implements the gate and protects against decoherence. Using numerical simulations, we can directly compare the performance of these two Hamiltonians for entanglement generation in the presence of pure dephasing.
\begin{figure}[H]
   \includegraphics{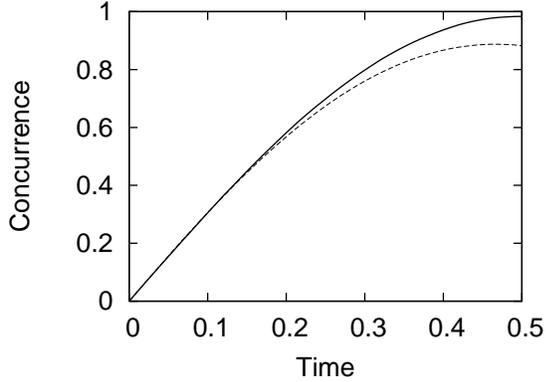}
   \centering
  	\caption{For the $\overline{\mbox{CNOT}}$ gate, behavior of entanglement versus time (in dimensionless units) using a bare Hamiltonian that only implements the gate (dashed line) and using control fields that both implement the gate and protect against all types of pure-dephasing (solid line) for the case of different baths. In the dimensionless units defined before, the parameters used are $n_1 = 2, n_2 = 1, \tau = 0.5, G = 0.03$.}
  	\label{CNOTdifferentbath}
\end{figure}
\begin{figure}[H]
   \includegraphics{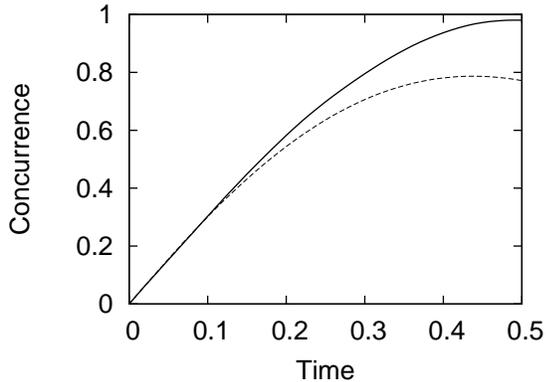}
   \centering
  	\caption{Performance of the $\overline{\mbox{CNOT}}$ gate in generating entanglement using a Hamiltonian that only implements the gate (dashed line) and using control fields that both implement the gate and protect against all types of pure-dephasing at the same time (solid line) for the common bath case. The parameters used are the same as for Fig.~\ref{CNOTdifferentbath}.}
  	\label{CNOTsamebath}
\end{figure}

Such a comparison is done in Figs. \ref{CNOTdifferentbath} and \ref{CNOTsamebath}.   The dashed curves depict the entanglement generation by the bare Hamiltonian
Eq.\,\eqref{CNOTsimplefield} and the solid lines
are for the performance by the control Hamiltonian given in Eq.\,\eqref{CNOTHamiltonian}.  As expected from an entangling gate, both Hamiltonians generate entanglement with similar performance in the beginning. However, after some time (for common-bath and different-bath cases), the effect of the environment is accumulated and eventually the bare Hamiltonian loses its battle against the environment, whereas for our constructed control Hamiltonian, the entanglement generation stays close to its expected value. By the time the gate operation is completed, much
better performance is achieved due to the application of continuous DD fields.

We now perform a similar task for the CZ gate in the presence of all types of decoherence.
The CZ gate is used to take a separable state to a fully entangled state. We compare the gate performance afforded by the control Hamiltonian given by Eq.\,\eqref{CZgateHamiltonian} with that of a bare Hamiltonian that solely implements the CZ gate.
The bare Hamiltonian, as already given by Eq.\,\eqref{CZsimplefield}, is
$$H_0 = \frac{\pi}{2\tau} \frac{1}{2}\left( \sigma_z^{(1)} + \sigma_z^{(2)} - \sigma_z^{(1)}\sigma_z^{(2)}\right).$$
Up to a global phase factor, this Hamiltonian implements the CZ gate in time $\tau$.  We stress that numerical simulations here
are no longer restricted to pure dephasing. Instead we are considering the most general case, allowing
errors such as uncorrelated bit flipping and dephasing as well as  `noisy' interaction between the two qubits.
\begin{figure}[H]
   \includegraphics{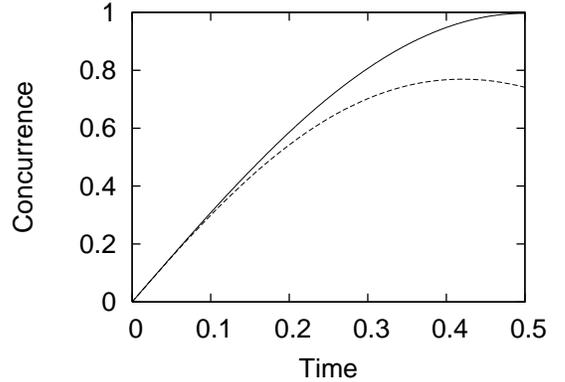}
   \centering
  	\caption{Evolution of entanglement using a bare Hamiltonian that only implements the gate (dashed line) and using control
  fields that both implement the gate and protect against decoherence at the same time (solid line) for the case of 15 different baths [Eq.\,\eqref{differentbathcoupling}]. The parameters (in dimensionless units defined before) are $\tau = 0.5, G = 0.02, \omega_c = \pi, n_x^{(1)} = 1, n_z^{(1)} = 2,  n_x^{(2)} = 4, n_z^{(2)} = 8$. The CZ gate here converts a separable state to an entangled state.}
  	\label{CZgatedifferentbath}
\end{figure}
In Figs. \ref{CZgatedifferentbath} and \ref{CZgatesamebath},  the dashed curves depict the performance of the bare Hamiltonian in Eq.\,\eqref{CZsimplefield}, whereas the solid curves show the performance of the control Hamiltonian we found in Eq.\,\eqref{CZgateHamiltonian}. Once again, the performance benefit is obvious. In both the different-bath and common-bath cases, with the continuous DD fields implemented, we are able to achieve almost perfectly entangled states even in the presence of all possible types of decoherence.  By contrast, the desired coherent evolution takes place with a clearly poor fidelity
if only a bare Hamiltonian is used.
\begin{figure}[H]
   \includegraphics{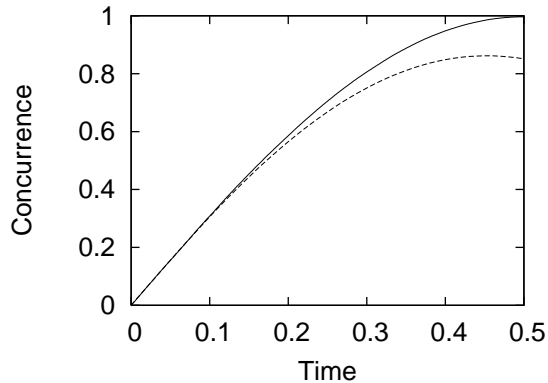}
   \centering
  	\caption{Considering the CZ gate, evolution of entanglement using a Hamiltonian that only implements the gate (dashed line) and using fields that both implement the gate and protect against decoherence at the same time (solid line) for the case of common bath [Eq.\,\eqref{samebathcoupling}]. The parameters used are the same as in Fig.~\ref{CZgatedifferentbath}.}
  	\label{CZgatesamebath}
\end{figure}

\section{CONCLUSION}
In this paper, we first asked and answered the following question: is it possible to use continuous fields to achieve (albeit low-order)
protection of two-qubit states as  a  universal dynamical decoupling approach?  By extending the methodology in Ref.~\cite{FanchiniPRA22007},
we have found a rather simple field configuration to achieve this task.  This associated decoherence control
is completely general in the sense that it is able to protect the state against all types of decoherence, so long as the frequency of the control fields is sufficiently large (with sufficient field strength) as compared with the environment cutoff frequency.
From a practical point of view, the very existence of a universal scheme is important if we do not have enough information about the environment.
The found continuous DD is also relatively simple - only local continuous and periodic fields are required.
 Our results are thus at least complementary to recent studies of universal pulsed DD for entanglement protection.  In particular,
 under the circumstances where multi-pulse DD is difficult to implement (e.g., due to the requirement of
 very small pulse intervals), then our universal continuous DD scheme provides one alternative.
 Furthermore, one can imagine using a combination of pulsed DD and continuous DD to reliably store quantum information.

We have also constructed continuous control fields to implement two-qubit gates in the presence of most general decoherence. This is important
for three reasons. First, it always takes a finite amount of time for a quantum gate to be implemented and as such a two-qubit gate must be protected against decoherence during the gate operation time. Second, during the implementation of a two-qubit gate, coherent evolution of the system itself complicates the issue of decoherence control due to the transformation between different types of quantum coherence properties. Third,
the implementation of two-qubit gates itself will unavoidably bring about noise in qubit-qubit interactions.
As seen from two case studies of universal two-qubit gate protection against both local and nonlocal noise, the required
control Hamiltonian is not too complicated, with the most involving component being oscillating qubit-qubit interaction terms.
Our treatment is  general in the sense that we have not considered any particular physical implementation of a two-qubit gate.
It would be interesting to apply our findings here to a particular physical realization of two-qubit gates. One excellent example would be
 in the recent implementation of  superconducting two-qubit gates using simple microwave fields \cite{ChowPRL2011}. In a second example from the trapped-ion context,
continuous microwave driving is already theoretically considered to protect two-qubit gates against noise due to magnetic field fluctuations and the thermal motion of the ions \cite{retzker}.
Of course, in such physical realizations, it may be the case that only a few noise sources contribute appreciably to decoherence and therefore
the required continuous DD fields may be simplified.
Our results here also lay a useful starting point for
future optimization studies~\cite{yupan,wilhelm}, by, for example,  first extracting some information about an environment.  Finally,
as pointed out earlier, this work offers an explicit and simple route to construct dynamically corrected two-qubit gates
 \cite{violaPRL2009,violaPRA2009} to fight against arbitrary system-environment coupling.

\begin{acknowledgements}
A.Z.C. would like to thank Jose-Garcia Palacios and Lee Chee Kong for helpful discussions on master equations, and Derek Ho and Juzar Thingna for insightful discussions on computational methods.  We are also grateful to Lorenza Viola for her very helpful comments on an early version of this manuscript.
\end{acknowledgements}

\appendix
\section{The Magnus Expansion}
The Magnus expansion \cite{Magnus1954} says that the unitary evolution operator, $U(t)$, corresponding to a time-dependent Hamiltonian, $H(t)$, can be written as
\begin{equation}
U(t) = \exp \sum_{i = 1}^\infty A_i(t)
\end{equation}
where
\begin{eqnarray}
A_1 = -i \int_0^t dt_1 H(t_1), \\
A_2 = -\frac{1}{2} \int_0^t dt_1 \int_0^{t_1} dt_2 [H(t_1),H(t_2)],
\end{eqnarray}
with higher order terms given by higher order commutator expressions. For further details, we refer the reader to Ref. \cite{Magnus2010}.

\section{Derivation of the master equation}
Here we present a conceptually simple derivation of the non-Markovian master equation that we have used. Such a master equation has been previously used in, for example, Ref.\,\cite{KurizkiPRL2004}. However, unlike Ref.\,\cite{KurizkiPRL2004}, we do not use any advanced techniques such as the projection-operator method to derive the master equation. Rather, we only use a simple perturbation theory. Note that it is essential that we do not make the Markov approximation since we are interested in decoherence control.

For simplicity, here we derive the master equation for a system interacting with an environment consisting of a single bath. The more general case of the environment consisting of multiple baths can be dealt with via a simple extension. As usual, we start off by writing down the Hamiltonian of the total system,
\begin{equation}
H_{\mbox{\scriptsize tot}} (t) = H(t) + V(t).
\end{equation}
$H(t) = H_{\text{S}}(t) + H_{\text{B}}$ is the Hamiltonian describing the free system and the bath, and, for notational convenience, here we use $V(t)$ instead of $H_{\text{SB}}$ to represent coupling between them. Note that these are Schrodinger picture operators. Any time dependence in the system Hamiltonian or the Hamiltonian describing the coupling is an explicit time dependence. This time dependence can arise, for instance, due to the application of external fields.

Consider system-environment operators of the form $ Y \otimes I_{\text{B}} $, where $Y$ is an operator acting on the Hilbert space of the system and $I_{\text{B}}$ denotes identity in the Hilbert space of the environment. If the state of the total system is described by the density matrix $\rho_{\mbox{\scriptsize tot}} (t) $, then the expectation value of the operator $Y$ is given by
\begin{equation}
\langle Y \rangle = \mbox{Tr}_{\text{S}} (Y \rho),
\end{equation}
where $\rho = \mbox{Tr}_{\text{B}} \rho_{\mbox{\scriptsize tot}}$. $\mbox{Tr}_{\text{B}}$ denotes taking trace over the bath degrees of freedom, while $\mbox{Tr}_{\text{S}}$ means tracing out the system degrees of freedom. Now, $\rho$ is our primary object of interest. We are trying to derive an equation of motion for $\rho$.
Note also that
\begin{equation}
\langle Y \rangle = \mbox{Tr}_{\text{S},\text{B}} [(Y \otimes I_{\text{B}}) \rho_{\mbox{\scriptsize tot}}].
\end{equation}
Now, for computational purposes, we express the density matrix in some basis, namely, $\rho_{mn}(t) = \bra{m} \rho(t) \ket{n} $. In particular,
 if we choose $Y = \ket{n}\bra{m} \equiv Y_{nm}$, we get
\begin{eqnarray}
\langle Y \rangle &=& \mbox{Tr}_{\text{S}} [Y_{nm} \rho(t)] \notag \\
&=& \bra{m} \rho(t) \ket{n} = \rho_{mn}(t) \notag \\
&=& \mbox{Tr}_{\text{S},\text{B}} [(Y_{nm} \otimes I_{\text{B}}) \rho_{\mbox{\scriptsize tot}}(t)] \notag \\
&=& \mbox{Tr}_{\text{S},\text{B}} [U^\dagger (t,t_0) (Y_{nm} \otimes I_{\text{B}}) U(t,t_0) \rho_{\mbox{\scriptsize tot}}(t_0)], \notag \\
\end{eqnarray}
where $U(t,t_0)$ is the unitary evolution operator describing the unitary evolution of the total system. The cyclic invariance property of the trace has also been used.
Defining $Y_{nm} \otimes I_{\text{B}} \equiv X_{nm} $, we observe that $U^\dagger (t,t_0) X_{nm} U(t,t_0)$ is just a Heisenberg picture operator. We refer to this operator as $X_{nm}(t)$, with the understanding that any $X$ operator with a time dependence is in the Heisenberg picture.
Therefore, we can write,
\begin{equation}
\rho_{mn}(t) = \mbox{Tr}_{\text{S},\text{B}} [\rho_{\mbox{\scriptsize tot}}(t_0) X_{nm}].
\end{equation}
It follows that
\begin{equation}
\label{rdmheisenbergeom}
\frac{d\rho_{mn}(t)}{dt} = \mbox{Tr}_{\text{S},\text{B}} \left[\rho_{\mbox{\scriptsize tot}}(t_0) \frac{dX_{nm}}{dt}\right].
\end{equation}

We now derive the Heisenberg equation of motion for a general system-environment operator, and substitute that in the above expression. After doing so, we take the trace over the system and the environment to obtain our master equation. We proceed with the derivation of this Heisenberg equation of motion using perturbation theory.

As in standard time-dependent perturbation theory, we set $U(t,t_0) = U_0(t,t_0) U_I(t,t_0) $ where $U_0(t,t_0) = U_{\text{S}}(t,t_0)U_{\text{B}}(t,t_0)$ describes the free evolution of the system and environment, and the $U_I(t,t_0)$ comes in due to the coupling with the environment. It follows then, that to first order in the system-environment coupling,

\begin{equation}
U(t,t_0) = U_0(t,t_0) \left(1 - i \int_{t_0}^t ds \tilde{V} (s) \right),
\end{equation}
where $\tilde{V}(s) = U_0^\dagger(s,t_0) V(s) U_0(s,t_0).$

We know that operators in the Heisenberg picture and operators in the Schrodinger picture are related by,
\begin{equation}
O_H(t) = U^\dagger (t,t_0) O(t) U(t,t_0),
\end{equation}
where $O(t)$ is a general Schrodinger picture operator, and $O_H(t)$ is its Heisenberg picture counterpart.

Considering $O_H(t)$ to be a general system-environment operator that has no explicit time dependence, we have the Heisenberg equation of motion,
\begin{equation}
\frac{dO_H(t)}{dt} = i[H_H(t),O_H(t)] + i[V_H(t),O_H(t)].
\end{equation}
Using our above perturbative expression for the unitary time evolution operator, and observing that $[V_H(t), O_H(t)]$ is a Heisenberg picture operator, we have,
\begin{align}
\label{heisenbergeom}
\frac{dO_H(t)}{dt} &= i[H_H(t),O_H(t)] + i[\tilde{V}(t),\tilde{O}(t)] \notag \\
&+ \int_{t_0}^t ds [[\tilde{V}(t),\tilde{O}(t)],\tilde{V}(s)].
\end{align}
We now set $O_H(t) = X_{nm}(t)$, and substitute Eq.~\eqref{heisenbergeom} in Eq.~\eqref{rdmheisenbergeom}. Each term of Eq.~\eqref{heisenbergeom} gives a term in the master equation, so we work them out one by one. We assume that the initial state is a product state $\rho(t_0) \otimes \rho_{\text{B}}$.

The first term of Eq.~\eqref{heisenbergeom}, i.e., $i[H_H(t),X_{nm}(t)]$, when substituted in Eq.~\eqref{rdmheisenbergeom}, leads to
\begin{eqnarray}
&& \mbox{Tr}_{\text{S},\text{B}} \lbrace(\rho(t_0) \otimes \rho_{\text{B}}) i [H_H(t),X_{nm}(t)]\rbrace \notag \\
 & =& i \sum_{m'n'} \Delta_{mnm'n'} \rho_{m'n'},
\end{eqnarray}
where $\Delta_{mnm'n'} \equiv \delta_{mm'} H_{n'n}(t) - \delta_{nn'} H_{mm'}(t).$
The second term of \eqref{heisenbergeom}, i.e., $i[\tilde{V}(t),\tilde{X}_{nm}(t)]$, leads to
$$ \mbox{Tr}_{\text{S},\text{B}} \lbrace i (\rho(t_0) \otimes \rho_{\text{B}}) U_0^\dagger (t,t_0) [V(t), X_{nm}] U_0(t,t_0)\rbrace. $$
The system-environment coupling is of the form,
\begin{equation}
 V(t) = F(t) \otimes B(t),
 \end{equation}
where $F(t)$ is an operator acting in the Hilbert space of the system and $B(t)$ is an operator acting in the Hilbert space of the environment. With such a coupling, we can then work out the trace over the environment. This is given by
\begin{equation}
\mbox{Tr}_{\text{B}} [\rho_{\text{B}} U_{\text{B}}^\dagger (t,t_0) B(t) U_{\text{B}}(t,t_0)] = \langle B(t) \rangle.
\end{equation}
We assume that this is zero. This is commonly referred to as `centering' of the bath.

The last term of \eqref{heisenbergeom}, when substituted in \eqref{rdmheisenbergeom}, gives us four terms. Here, we only show the working for one of them.
The rest can be worked out in almost the same way. We now proceed to simplify
\begin{equation}
\label{tracesystembath}
 \mbox{Tr}_{\text{S},\text{B}} [(\rho(t_0) \otimes \rho_{\text{B}}) \int_{t_0}^t ds\ \tilde{V}(t)\tilde{X}_{nm}(t)\tilde{V}(s)].
\end{equation}
The trace over the environment gives
\begin{eqnarray}
&&\mbox{Tr}_{\text{B}} [\rho_{\text{B}} U_{\text{B}}^\dagger (t,t_0) B(t) U_{\text{B}}(t,t_0) U_{\text{B}}^\dagger(s,t_0) B(s) U_{\text{B}}(s,t_0)] \notag \\
&=& \langle \tilde{B}(t) \tilde{B}(s) \rangle = C_{ts}.
\end{eqnarray}
The trace over the system gives
$$ \mbox{Tr}_{\text{S}} [\tilde{\rho} (t) F(t) Y_{nm} U_{\text{S}} (t,s) F(s) U_{\text{S}}^\dagger (t,s)]. $$
We can now use the completeness relations of the basis states to write
\begin{equation}
F(t) = \sum_{n'm'} F_{n'm'}(t) Y_{n'm'},
\end{equation}
\begin{equation}
F(s) = \sum_{\alpha \beta} F_{\alpha \beta} (s) Y_{\alpha \beta},
\end{equation}
\begin{equation}
U_{\text{S}}(t,s) = \sum_{\mu \nu} U_{\text{S}}^{\mu \nu} (t,s) Y_{\mu \nu},
\end{equation}
\begin{equation}
U_{\text{S}}^\dagger (t,s) = \sum_{\mu' \nu'} U_{\text{S}}^{\dagger \mu' \nu'} Y_{\mu' \nu'} = \sum_{\mu' \nu'} U_{\text{S}}^{* \nu' \mu'} Y_{\mu' \nu'}.
\end{equation}
Substituting these relations back, and assuming that our basis states are orthonormal, we can simplify the trace over the system to
$$ \sum_{n' m'} \sum_{\mu' \nu'} F_{n'n}(t) F_{m'\mu'}(s) U_S^{mm'}(t,s) U_S^{\dagger \mu' \nu'}(t,s) \rho_{\nu' n'}(t), $$
where $\tilde{\rho}(t)$ has been replaced by $\rho(t)$. This is justified since the correction gives us terms of higher order in the coupling strength in the master equation.
For convenience, we define,
\begin{eqnarray}
g^{m\nu'}_{m' \mu'} (t,s) \equiv U_{\text{S}}^{mm'}(t,s) U_{\text{S}}^{* \nu' \mu'} (t,s).\\ \nonumber
\end{eqnarray}
Putting all the above expressions together, Eq.\,\eqref{tracesystembath} simplifies to
\begin{equation}
\int_{t_0}^t ds \sum_{n'm'} \sum_{\mu \nu} F_{n'n}(t) F_{m' \mu}(s) g_{m'\mu}^{m\nu}(t,s) \rho_{\nu n'}(t) \langle \tilde{B}(t) \tilde{B}(s) \rangle.
\end{equation}
After working out all other terms,  we finally arrive at the master equation

\begin{align}
\frac{d\rho_{mn}(t)}{dt} &= i \sum_{m'n'} \Delta_{mnm'n'}(t) \rho_{m'n'}(t) \quad +  \quad \sum_{m'n'} \rho_{m'n'}(t) \notag \\
&\times \biggl\lbrace  \int_{t_0}^t ds \sum_{\mu \nu} \sum_l \Bigl[ \delta_{\nu m'} F_{n'n}(t) F_{l \mu} (s) g^{m\nu}_{l \mu} (t,s) C_{ts}\notag \\
&+\ \delta_{\nu n'} F_{mm'}(t) F_{\mu l} (s) g^{\nu n}_{\mu l} (t,s) C_{st}  \notag \\
&-\ \sum_{l'} ( \delta_{mm'} \delta_{\mu n'} F_{l'n}(t) F_{\nu l} (s) g^{\mu l'}_{\nu l} (t,s) C_{st}   \notag \\
&+\ \delta_{\mu m'} \delta_{n n'} F_{ml'}(t) F_{l \nu} (s) g^{l' \mu}_{l \nu} (t,s) C_{ts}) \Bigr]\biggr\rbrace.
  \ \
\end{align}

From a computational point of view, this is the most useful form of the master equation. However, it is also useful to see the basis independent form of the master equation. Using the completeness relations of the basis states in the master equation above, one can remove the summations. After taking into account the possibility of multiple baths, one then ends up with the basis independent form of the master equation, which is given by Eq.\,\eqref{masterequation} in the main text.

In order to solve the master equation, we first note that the system Hamiltonian is generally explicitly time-dependent. In such a case, we cannot, in general, calculate $U_{\text{S}}(t,s)$ in a simple way. Furthermore, solving the master equation itself becomes much more involved because of this explicit time-dependence, since the integrand in the master equation now depends explicitly on $t$. For our purpose, however, since we are only interested in short times, we can still use a straightforward method to solve the master equation. Even though we know $U_{\text{S}}(t,s)$ for our case, we choose to start from the Hamiltonians - this serves as a check that the Hamiltonians have been calculated correctly. We first use the split-operator method \cite{Feit1982,Kouri1990,Kouri1991} to calculate $U_{\text{S}}(t,s)$, then perform the integration in the master equation numerically, and we then finally solve the differential equation using the fourth-order Runge-Kutta (RK4) algorithm \cite{Rabinowitzbook}. We also note that there are considerably more involved and more efficient methods to solve such driven open system problems based on a decomposition of the bath spectral density (see, for example, Ref. \cite{KartoffelJCP2004}).

\end{document}